\documentclass[11pt]{article}

\usepackage{amsthm,amsmath,amssymb,bbm,bm}
\usepackage{multirow}
\usepackage[pdftex]{graphicx}
\usepackage{subfigure}
\usepackage{wrapfig}
\usepackage{CJK}
\usepackage{array}
\usepackage{url}
\usepackage{booktabs}
\usepackage{algorithm}
\usepackage{algorithmic}
\usepackage{threeparttable}
\usepackage{placeins}
\usepackage{mathtools}
\usepackage{romannum}
\usepackage{mathrsfs}
\usepackage{dsfont}
\usepackage{natbib}
\usepackage{relsize}
\usepackage{rotating}
\usepackage{enumitem}
\usepackage{etoc}
\usepackage{setspace}

\usepackage[usenames,dvipsnames,svgnames,table]{xcolor}

\usepackage[colorlinks=true,
            linkcolor=red,
            anchorcolor=blue,
            citecolor=blue,
            urlcolor=blue]{hyperref}

\makeatletter
\newcommand{\vast}{\bBigg@{3}}
\makeatother

\usepackage{chngcntr}
\counterwithin{equation}{section}

\theoremstyle{definition}

\newtheorem{thm}{Theorem}

\counterwithin{thm}{section}
\newtheorem{rmk}{Remark}[thm]
\counterwithin{rmk}{section}

\def\##1\#{\begin{align}#1\end{align}}
\def\$#1\${\begin{align*}#1\end{align*}}

\newtheorem{assump}{Assumption}

\makeatletter
\newcommand*{\rom}[1]{\expandafter\@slowromancap\romannumeral #1@}
\makeatother

\def\beq#1\eeq{\begin{equation}#1\end{equation}}
\def\baa#1\eaa{\begin{eqnarray}#1\end{eqnarray}}
\def\bal#1\eal{\begin{align}#1\end{align}}

\def\T{\text{T}}

\usepackage{geometry}
\begin{document}

\pagenumbering{arabic}

\title{ {\LARGE A Parsimonious Personalized Dose Finding Model via Dimension Reduction} }

\author{Wenzhuo Zhou\thanks{Department of Statistics, University of Illinois Urbana-Champaign, Champaign, IL 61820; email: \href{mailto:wenzhuo3@illinois.edu}{wenzhuo3@illinois.edu}.}, \,
Ruoqing Zhu\thanks{Department of Statistics, University of Illinois Urbana-Champaign, Champaign, IL 61820; email: \href{mailto:rqzhu@illinois.edu}{rqzhu@illinois.edu}.} \,
and Donglin Zeng\thanks{Department of Biostatistics, University of North Carolina at Chapel Hill, Chapel Hill, NC 27599; email: \href{mailto:dzeng@bios.unc.edu}{dzeng@bios.unc.edu}.}
}
\date{}
\maketitle

\begin{abstract}
Learning an individualized dose rule in personalized medicine is a challenging statistical problem. Existing methods often suffer from the curse of dimensionality, especially when the decision function is estimated nonparametrically. To tackle this problem, we propose a dimension reduction framework that effectively reduces the estimation to a lower-dimensional subspace of the covariates. We exploit that the individualized dose rule can be defined in a subspace spanned by a few linear combinations of the covariates, leading to a more parsimonious model. The proposed framework does not require the inverse probability of the propensity score under observational studies due to a direct maximization of the value function. This distinguishes us from the outcome weighted learning framework, which also solves decision rules directly. Under the same framework, we further propose a pseudo-direct learning approach that focuses more on estimating the dimensionality-reduced subspace of the treatment outcome. Parameters in both approaches can be estimated efficiently using an orthogonality constrained optimization algorithm on the Stiefel manifold. Under mild regularity assumptions, the results on the asymptotic normality of the proposed estimators are established, respectively. We also derive the consistency and convergence rate for the value function under the estimated optimal dose rule. We evaluate the performance of the proposed approaches through extensive simulation studies and a warfarin pharmacogenetic dataset.
\end{abstract}

\noindent {\bf keywords}
Individualized Dose Rule; Dimension Reduction; Semiparametric Inference; Direct Learning; Propensity Score; Stiefel Manifold.

\section{Introduction}\label{sec:intro}

Personalized medicine is a medical procedure that aims to improve an individual patient's health outcome by a tailored medical treatment based on the patient's genetic, prognostic, and clinical information. It has received much attention from the statistical and clinical research communities due to patients' heterogeneity and the difficulties involved in implementations. There is an extensive literature on developing statistical methodologies for estimating individualized treatment rules, in both single and multiple decision points. Examples include risk level calibration \citep{cai2010calibrating}, penalized models \citep{lu2013variable, tian2014simple}, machine learning approaches \citep{foster2011subgroup, kang2014combining, laber2015tree, loh2015regression}, two-step regression-based approaches \citep{zhao2009reinforcement, qian2011performance, moodie2014q, zhu2016individualizing} and many others. The readers may refer to \cite{kosorok2019precision} for a comprehensive review.

A popular framework for estimating the optimal individualized treatment rule is outcome weighted \citep{zhao2012estimating}. The advantage of this framework is directly estimating the treatment decision without modeling the relationship between the treatment and outcome. \cite{zhao2012estimating} developed an approach for binary treatments. It has been extended to other settings such as backward and simultaneous outcome weighted learning \citep{zhao2015new}, the tree-based method of \cite{zhu2017greedy}, residual weighted learning \citep{zhou2017residual}, and augmented outcome weighted learning \citep{liu2018augmented}. For multicategory treatment settings, existing approaches include the offset tree algorithm \citep{beygelzimer2009offset}, contrast weighting \citep{tao2017adaptive}, and others \citep{zhang2018multicategory, zhou2018outcome, huang2019multicategory}. Extensions to dose-finding \citep{chen2016personalized} are statistically more challenging due to the cardinality of treatment options.

To obtain more flexible decision rules, many existing methods employ machine learning or nonparametric approaches. Support vector machines \citep{cortes1995support} and random forests \citep{breiman2001random} are extensively used in the personalized medicine literature. However, as the complexity of decision rules increases, the sheer number of covariates makes it difficult to estimate the underlying model accurately. Besides linear models, a promising approach is sufficient dimension reduction \citep{cook2009regression} of the feature space. In a classical setting, sufficient dimension reduction assumes that the response relies only on several linear combinations of the covariates, which greatly reduces the feature space. Furthermore, dimension reduction allows for better interpretability of the fitted model, which is particularly appealing in personalized medicine. Existing works of sufficient dimension reduction include methods proposed by \cite{li1991sliced, cook1991comment,xia2002adaptive,ma2012semiparametric,ma2013efficient} and many others. We refer readers to \cite{adragni2009sufficient} and \cite{ma2013review} for more details of this topic.

In this paper, we propose a dimension reduction framework for estimating the optimal individualized dose rule. An immediate advantage is that the nonparametrically optimal dose rule is adaptive to a low-dimensional covariate space. This circumvents the curse of dimensionality and allows for better interpretability. We further propose two approaches: a direct learning approach and a pseudo-direct learning approach. In the direct learning approach, we exploit that the dose rule, instead of the outcome, can be modeled in a linear subspace of the covariates. This leads to a key advantage that we can directly estimate the optimal dose rule without modeling propensity scores in an observational study. On the other hand, the pseudo-direct learning approach is achieved in two-stages. In the first stage, we estimate the low-dimensional subspace of the outcome, while in the second stage, the optimal dose rule can be estimated on the dimensionality-reduced subspace. Interestingly, the pseudo-direct learning approach has a close connection with the partial dimension reduction framework \citep{feng2013partial} when we regard the dose treatment as a conditional variable. The proposed estimators are constructed under a semiparametric framework, where we utilize an orthogonality-constrained optimization algorithm to solve for the parameters.

\section{Proposed Method}\label{sec:method}

\subsection{Personalized Dose-finding}

Consider a dose assignment $A \in \mathcal{A} = [0, 1]$, where $\mathcal{A}$ is a safe dose range. Let $X = (X_{1}, X_{2}, ..., X_{p})^\T \in \mathcal{X}$ denote the prognostic covariates, where $\mathcal{X}$ is the feature space. Let $R$ be the observed clinical outcome. Without loss of generality, we assume that a larger value of $R$ is more desirable. An individualized treatment rule  $f$, is a map from $\mathcal{X}$ to a dose assignment space $\mathcal{A}$. To properly define an optimal individualized treatment rule $f_{\text{opt}}(X)$, we consider a potential outcomes framework \citep{rubin1974estimating,robins1986new}. Let $R^{*}(a)$ be the potential outcome under a given dose level $a$, and we make the following common and well-studied assumptions: (i) strong ignorability \citep{robins1986new}, $A \perp \!\!\! \perp R^{*}(a) \mid X$, for all $a \in \mathcal{A}$; (ii) consistency, $R=R^{*}(A)$; (iii) positivity, $P(A = a \mid X=x) > 0$, almost surely. Under a randomized dose trial, the strong ignorability assumption is trivially satisfied; however in an observational study, it could be unverifiable \citep{bang2005doubly}. Also, the assumption (iii) can be relaxed under our proposed dimension reduction framework and a corresponding discussion will be presented in Remark \ref{positive}. Under the above assumptions, we are able to estimate the optimal treatment rule by using the value function proposed in \cite{qian2011performance}:
\#
V_f = E^f(R) = \int R \, d P^f,
\label{eq:orvalue}
\#
where $E^f$ is the expectation with respect to $P^f$ which is defined as the restricted joint distribution of $\{X, R, A = f(X)\}$. Then the optimal individual treatment rule is the maximizer of the value function, that is $f_{\text{opt}}(X) = \operatorname*{argmax}_{f} V_{f} $. When $\mathcal{A}$ is a binary space such that $\mathcal{A}= \{-1,1\}$, \cite{zhao2012estimating} estimate $f_{\text{opt}}(X)$ by using the value function in \eqref{eq:orvalue}. For dose-finding problems, \cite{chen2016personalized} extend the binary outcome weighted learning framework to handle continuous treatment options by using a local approximation of the value function:
\# \begin{split}
V_{f} &=  \lim_{\phi \to 0^+}  E\Bigg(\frac{R \mathbbm{1}\big[ A\in \big\{ f(X) - \phi, f(X) + \phi \big \} \big] }{2\phi P(A \mid X)} \Bigg) \doteq \lim_{\phi \to 0^+}   {\widetilde{V}}_{f,\phi},
\label{eq:limitval}
\end{split}
\#
where $P(A\mid X)$ is the randomization probability of $A$ given $X$. The dose rule can then be solved using the sample version of this approximation.

\subsection{Dimensionality-Reduced Personalized Dose Rule}

In practice, patients' prognostic, genetic, and clinical information usually consists of a large number of covariates. This often creates difficulties for estimating the optimal dose rule. For example, support vector regressions used in the outcome weighted learning approaches can be sensitive to the dimension of covariates \citep{dasgupta2013feature}. It is then desirable to construct an optimal dose rule in a dimensionality-reduced space. In particular, if there is a $p \times d$ constant matrix $B$ with $d < p$, such that the optimal dose rule can be re-defined as
\#
f_{\text{opt}}(X)=f_{\text{opt}}(B^\T X),
\label{eq:drrule}
\#
then its estimation can be more efficient. This structure can not only capture the majority of the information carried by the original features but also enjoy better interpretability and can be easier to implement. This falls into the sufficient dimension reduction framework \citep{li1991sliced, cook2009regression}, which is a long-standing and promising technique. With a slight abuse of notation, let $f(B^\T X): \mathbb{R}^d \rightarrow \cal{A}$ be a dose rule as an analogy of $f(X)$ in the dimensionality-reduced space. We will focus on two goals: estimating the dimension reduction basis matrix $B$ and the optimal rule $f_\text{opt}$ in the dimensionality-reduced space.

\subsection{The Direct Learning Approach}\label{sec:dirct}

Let us define a new version of the value function as follows:
\begin{align}
V_f(B) &= E\left[ E \{ R \mid X, A = f(B^\T X) \} \right] \nonumber \\
&= E\left[ E \{ R \mid B^\T X, A = f(B^\T X) \} \right].
\label{eq:drvalue}
\end{align}
In this equation, the assigned dose level is a deterministic quantity as a function of $B^T X$. Also, $V_f(B)$ is not necessarily the same as $V_f$ since $f(B^\T X)$ is more restrictive than $f(X)$. However, we can see that the two attain the same maximum under the assumption of the previously defined dimension reduction optimal dose rule \eqref{eq:drrule}, that is
\begin{align}
\max_{f} V_f = \max_{f, B} V_f(B). \nonumber
\end{align}

The formulation in \eqref{eq:drvalue} suggests taking expectation on the restricted distribution $\{R, B^\T X, A = f(B^\T X)\}$ while searching for the best direction $B$ to directly maximize the value function. This is similar to a regression framework. However, it does not assume that the outcome $R$ depends only on $B^\T X$. Instead, $E \{ R \mid B^\T X, A = f(B^\T X) \}$ can be regarded as a sub-population average who has the same suggested dose level $f(B^\T X)$ with covariate $B^\T X$. The second equality in \eqref{eq:drvalue} suggests that we do not need to model the regression outcome using the full information of $X$. This strategy distinguishes us from the outcome weighted learning framework \citep{zhao2012estimating}, which utilizes the Radon-Nikodym theorem and estimates the value function on the distribution of $(R, X, A)$. Interestingly, this also allows us to avoid estimating the propensity score in an observational study setting. Another advantage of our proposal is that $B^\T X$ is a low-dimensional vector which circumvents the curse of dimensionality. This leads to the following penalized sample version of the value function through kernel approximations of the conditional expectation $E\{R \mid B^\T X, A = f(B^\T X)\}$ for each subject:
\#
V_{n, f}(B) = \frac{1}{n}\sum\limits_{j=1}^n\frac{ \sum_{i=1}^n R_i K_{v}\Big[ \big\{X_j^\T B, f(B^\T X_j)\big\}^{\T} -\big( X_i^\T B, A_i\big)^\T\Big]}{\sum_{i=1}^n K_{v}\Big[ \big\{ X_j^\T B,f(B^\T X_j)\big\}^{\T} -\big(X_i^\T B, A_i\big)^\T\Big]} - \lambda_n \| f \|^2. \label{eq:sampleval}
\#
Here, $K_{v}(u) = \prod_{k=1}^{d+1}K(u_{k}/h)/h$ is a $d+1$ dimensional kernel function with a $d+1$-dimensional bandwidth vector $v = (h, ..., h)^\T$. $K(u)$ is a univariate kernel and $d$ is the true structural dimension. Lastly, $\|f\|$ is some seminorm for $f$ and $\lambda_n$ serves as a penalty on the complexity of $f$. The optimization problem is then defined as
\#
( \widehat B,\ \widehat f_{\text{opt}}  )= \operatorname*{argmax}_{B \in {\mathbb{R}}^{p \times d},\, f} V_{n,f}(B).
\label{eq:optim}
\#
Maximizing objective function is still a challenging task because of the two unknowns, $f$ and $B$. We consider an update scheme which alternates between $B$ and $f$, motivated by a series of semiparametric dimension reduction models \citep{ma2012semiparametric, ma2013efficient, sun2019counting}. To be specific, maximizing $B$ over $V_{n, f}(B)$ can be viewed as searching for the dimension reduction space, which can be carried out using the computational approach proposed by \cite{sun2019counting}. On the other hand, when fixing a matrix $B$, solving $f$ reduces to a personalized dose-finding problem on the reduced space $B^\T X$. Details of the proposed algorithm is given in Section \ref{sec:alg}. Furthermore, we shall see from Section \ref{sec:theory} that the numerical optimization procedure may affect the asymptotic properties of the estimator. Hence, we will consider two different strategies when optimizing \eqref{eq:optim}, one of which involves a sample-splitting procedure, detailed in Section \ref{sec:alg}.

\begin{rmk}\label{positive}
Existing propensity score based approaches usually require the positivity assumption such that $P(A = a \mid X = x) > 0$. This is not needed for our approach. Instead, we require that $P(A = a \mid B^\T X = z) > 0$ for all $B$, $z$ and $a$. This is a similar but slightly weaker assumption to ensure that nonparametric estimation can be performed on $(B^\T X, A)$. Such the positivity assumption is natural in nonparametric regressions \citep{ichimura1991semiparametric, andrews1995nonparametric}.
\end{rmk}

\subsection{The Pseudo-Direct Learning Approach}\label{sec:pdirct}

We may also make a slightly stronger assumption on the outcome to impose a multiple-index model. This connects the proposed framework with a partial dimension reduction model \citep{feng2013partial}. In particular, let
\#
E(R \mid X, A) = M(B^\T X, A),
\label{eq:index}
\#
where $M(\cdot)$ is an unknown link function. The implication of $B$ is slightly different than the direct learning approach. Here, we assume that the outcome $R$ relies completely on $B^\T X$ and $A$, whereas in the direct learning approach, $R$ may still depend on the entire covariates $X$. However, it is easy to see that the optimal treatment $f_{\text{opt}} = \operatorname*{argmax}_{a} M(B^\T X, A=a)$
must still be a function of $B^\T X$. If we estimate $B$ matrix first, the does rule can be estimated in the reduced covariate space $B^\T X$. Therefore, we consider a two-stage approach: In the first stage, we identify the dimensionality-reduced space by estimating $B$ matrix in \eqref{eq:index}; In the second stage, we apply an existing dose-finding approach to learn the optimal dose rule using $B^\T X$. In the first step, the conditional mean function can be estimated by
\#
\widehat M\big( B^\T x, a \big) = \frac{ \sum_{i=1}^n R_i K_{v}\big\{\big({X^\T_i B},A_i\big)^\T -\big({x^\T B},a\big)^\T\big\}} {\sum_{i=1}^n K_{v}\big\{\big({X^\T_i B},A_i\big)^\T  - \big({x^\T B},a\big)^\T \big\}},\label{eq:estindex}
\#
where the multidimensional kernel $K_{v}(\cdot)$ follows our previous definition in \eqref{eq:sampleval}. The matrix $B$ can be solved by minimizing the least-squares objective function
\#
\psi({B}) = \frac{1}{n}\sum_{i=1}^n \big\{ R_i - \widehat M\big(B^\T X_i, A_i \big) \big\}^2.
\label{eq:indobj}
\#
Hence, the pseudo-direct learning approach is to solve the minimization problem:
\#
\widehat {B} = \operatorname*{argmin}_{B \in {\mathbb{R}}^{p \times d} } \,\,\psi({B}).
\label{eq:indexoptim}
\#
The implementation of pseudo-direct learning will be discussed in Section \ref{sec:pseudo-alg}. Note again that there is a connection between our approach and the partial dimension reduction problem \citep{feng2013partial}, which is done by an estimation matrix and eigen-decomposition. Comparisons will be provided in Supplementary Material.

\section{Implementation and Algorithms}\label{sec:alg}

\subsection{Direct Learning Algorithm}

We first discuss a general procedure for updating $f$ and $B$. However, our theoretical investigation suggests that, a naive optimization can only guarantee consistency of the estimator. To achieve asymptotic normality of $\widehat{B}$, a sample-splitting procedure \citep{bickel1982adaptive,powell1989semiparametric} can be incorporated. The sample-splitting procedure will be described in Section \ref{sec:alg:split}, while in the current section, we focus on a standard procedure. Consider updating $f$ when given $B$ at the $t$-th iteration. We denote $B^{(t)}$ as the current value of the basis matrix $B$, then we turn to solve:
\#
\operatorname*{argmax}_{f} \, {V_{n,\, f}} \big(B^{(t)}\big).
\label{eq:foptim}
\#
This is to find an individualized dose rule in the dimensionality-reduced subspace spanned by $B^{(t)}$. This can be done by considering $f$ from the Reproducing kernel Hilbert Space $\cal H$:
\#
f^{(t)}(\cdot) =& \sum^n_{j=1} w_j K(\cdot, {B^{(t)}}^\T X_j), \label{eq:fupdate:simple}
\#
and the penalty term becomes $\lVert w \rVert^2$. Hence, plugging \eqref{eq:fupdate:simple} into the optimization problem \eqref{eq:foptim}, we solve for the parameter vector $w = (w_1, \ldots, w_n)^T$. This can be done by performing gradient descent. The gradient descent solution may be influenced by its initial value. In practice, we can obtain a warm start $w$ using a kernel ridge regression. Details of this implementation are provided in Supplementary Material.

We then proceed to solve $B$ by fixing $f$. As pointed out by \cite{sun2019counting}, the difficulty lies in guaranteeing the column rank of $B$ when all elements of $B$ are of free-changing parameters. Besides, any scale change of an entire column in $B$ gives essentially the same solution for estimating the optimal rule. Therefore, we consider a fixed $d$ and estimate $B$ in a restricted space such that $B^\T B = {I}_{d \times d}$, which leads to
\#
\widehat B= \operatorname*{argmax}_{B \in {\mathbb{R}}^{p \times d}, \, B^\T B = I } V_{n,\widehat f}(B).
\label{eq:boptim}
\#
This orthogonality constrained space is commonly known as the Stiefel manifold. We use a first-order updating procedure proposed by \cite{wen2013feasible}, which preserves the solution in the manifold. The procedure works by calculating the gradient matrix $G$ and the skew-symmetric matrix $Q$ using
\#
G = \frac{-\partial V_{n, f^{(t)}}(B)}{\partial B} \Big\rvert_{B^{(t)}} \,\, \text{and} \,\, Q = G {B^{(t)}}^\T - {B^{(t)}} G^\T,
 \label{eq:grad}
\#
respectively. Here, the gradient matrix $G$ can be approximated numerically. Then, we update
\#
B^{(t+1)} = {\bigg(I + \frac{\tau}{2}Q\bigg)}^{-1}\bigg(I - \frac{\tau}{2}Q\bigg)B^{(t)},
\label{eq:update}
\#
where $\tau$ is a small step size chosen to satisfy the Armijo-Wolfe conditions at the current iteration. This updated $B$ enjoys the property that ${B^{(t+1)}}^\T B^{(t+1)} = {B^{(t)}}^\T B^{(t)} = I$, which preserves the orthogonality. Now, this completes the update of $B$ when fixing $f$. A summary of the  estimating procedure is provided in Algorithm \ref{alg:direct}.

\begin{algorithm}\label{alg:direct}
Direct learning alternating update algorithm.


\begin{tabbing}
 \qquad \enspace {Initialize}: ${\cal D} = \{X_i, A_i, R_i\}_{i=1}^n$, $B^{(0)}$ such that ${B^{(0)}}^\T B^{(0)} = I$, $\varepsilon \leftarrow 10^{-8}$. \\
\qquad \enspace {For} $t = 1$ to $t =$ max.iter:\\
\qquad \qquad Fixing $B^{(t)}$, solve the optimization problem \eqref{eq:foptim} by updating $w$ in \eqref{eq:fupdate:simple}.\\
\qquad \qquad Fixing $f^{(t+1)}(\cdot)$, numerically approximate $G$ and $Q$ based on equation \eqref{eq:grad}.\\
\qquad \qquad Use a line search algorithm to find a step size $\tau$ in \eqref{eq:update}.\\
\qquad \qquad Update $B^{(t+1)} \leftarrow {\big(I + \frac{\tau}{2}Q\big)}^{-1}\big(I - \frac{\tau}{2}Q\big)B^{(t)}.$\\
\qquad \qquad Stop if $ \lVert G \rVert \leq \varepsilon$.\\
\qquad \enspace {Return}: $\widehat B = B^{(t+1)}$ and $\widehat f_{\text{opt}} = f^{(t+1)}$.
\end{tabbing}
\end{algorithm}

\subsection{Direct Learning with Sample-splitting}\label{sec:alg:split}

In Algorithm \ref{alg:direct}, we utilize the whole sample data when updating $B$ and $f$. Our theoretical analysis shows that $\widehat B$ based on this algorithm is a consistent estimator, however, it may not achieve asymptotical normality. To address this issue, we incorporate a sample-splitting strategy \citep{bickel1982adaptive,powell1989semiparametric}. Specifically, we randomly split the whole sample data into two subsets, ${\cal D}_1 = \{X_i, A_i, R_i\}_{i=1}^{n_1}$ and ${\cal D}_2 = \{X_j, A_j, R_j\}_{i=n_1+1}^{n}$ using, for example, $n_1 = n/2$. When fixing $f$ and updating $B$, we use ${\cal D}_1$; when fixing $B$ and updating $f$, we use ${\cal D}_2$. The detailed algorithm is provided in Supplementary Material.

\subsection{Pseudo-Direct Learning Algorithm}\label{sec:pseudo-alg}
The pseudo-direct learning method concerns solving the optimization problem defined in \eqref{eq:indexoptim}. Similar to the direct learning approach, we solve this in the Stiefel manifold:
\#
\widehat {B} = \operatorname*{argmin}_{B \in {\mathbb{R}}^{p \times d}, \, B^\T B = I } \,\, \psi({B}).
\label{eq:indexoptim2}
\#
where $\psi({B})$ is the $\ell_2$ loss objective function defined in \eqref{eq:indobj}. The same updating scheme in \eqref{eq:update} can be used to obtain the solution. We omit the details of the algorithm here since the procedure is largely identical to updating of $B$ when fixing $f$ in the direct learning approach, except that this is a minimization instead of maximization.

\begin{rmk}
In both the direct, including direct-split, and pseudo-direct learning algorithms, we use the Gaussian kernel function with a bandwidth $h = \{4/(d+2)\}^{1/(d+4)}n^{-1/(d+4)}\widehat\sigma$ on each dimension, where $\widehat\sigma$ is the estimated standard deviation of the corresponding variable.
\end{rmk}

\begin{rmk}
In all algorithms, we suggest an initial value of $B^{(0)}$ obtained from the partial-SAVE proposed in \citep{feng2013partial}, which is computationally fast. Although the method in \cite{feng2013partial} is proposed for the partial linear multiple index model, it can still serve as a good warm start based on our numerical experience.
\end{rmk}

\section{Theoretical Properties}\label{sec:theory}

In this section, we investigate the theoretical properties of the proposed estimators and algorithms. In Theorem \ref{thm:psudo}, we show that $\widehat{B}$ obtained from the pseudo direct learning approach is asymptotically normal. Then, we analyze the consistency of the direct learning estimator in Theorem \ref{thm:direct:consist}. This result also applies to a direct-split algorithm with slight modifications. In Theorem \ref{thm:split:normal}, we present the asymptotic normality of the estimator obtained from the direct-split algorithm. Lastly, the convergence rate of the value function is established in Theorem \ref{thm:value}.

To facilitate later arguments, we consider an upper-block diagonal version of the basis matrix $B$ following the idea in \citep{ma2013efficient}. This is mainly for the identifiability concern. To be specific, we can always find a rotation matrix $U$ such that $BU = (I_{d}, B^\T _{l})^{\T}$ where $d$ is the true structural dimension, $I_{d}$ is a $d \times d$ identity matrix, and $B_{l}$ is a $(p-d) \times d$ matrix. Hence, the basis matrix $B$ is identifiable by solving elements in $B_{l}$. We further define the concatenation of the columns in any arbitrary $p \times d$ parameterized matrix $B$ as $\text{vecl}(B) =  \text{vec}({B_l}) =(B_{d+1,1},...,B_{p,1},...,B_{d+1,d},...,B_{p,d})^\T$. We also denote $B_0$ as the true basis matrix after this parameterization. All proofs in this section are provided in Supplementary Material.

We first provide several regularity assumptions required for showing the asymptotic normality of the pseudo-direct learning estimator. Assumption \ref{assump1} states some smoothness conditions with respect to the underlying conditional mean function and density function for the convergence of the kernel estimator. Assumption \ref{assump2} ensures the information matrix of $B_0$ is non-singular. Assumption \ref{assump3} provides bandwidth conditions for the kernel estimator.

\begin{assump} \label{assump1}
Let $\widetilde{X} = (\check{X}^\T,...,\check{X}^\T) \in {\mathbb{R}}^{(p-d)\times d}$ with $\check{X} = (X_{d+1},...,X_p)^\T$ and $Z= {B}^\T X$ for given matrix $B$. We denote $(\cdot)^{\otimes}$ as the Kronecker power of a vector, and $p(z, a)$ as the probability density function of $(Z,A)$. For $k=1,2$ and $\alpha \in \mathbbm{Z}^{+}$, we denote ${\alpha}$-th partial derivatives of ${M}(z,a)$, $E\{(\widetilde{X} - \widetilde{x})^{\otimes k} \mid Z= z \}$ and $p(z,a)$ with respect to the argument $z$ as $\partial^{\alpha}_{z}{M}(z,a)$, $\partial^{\alpha}_{z}E\{(\widetilde{X} - \widetilde{x})^{\otimes k} \mid Z= z \}$ and $\partial^{\alpha}_{z} p(z,a)$, respectively. These derivatives are Lipschitz continuous over $(z,a)$ with the Lipschitz constant independent of $(z, a)$.
\end{assump}

\begin{assump} \label{assump2}
For $l = 0, 1$, let
\$
{M}^{[1]}_{l}({{B}^\T_0} x,a) &= \partial_{\text{vecl}({{B}})}[E\{(\widetilde{X} - \widetilde{x})\mid {{B}^\T_0} X ={{B}^\T_0} x\} \{{M}({{B}^\T_0} x,a)\}^l p({{B}^\T_0}x,a)] \\
\text{and} \,\,\, \,\, {M}^{[1]}({{B}^\T_0} x,a) &= \sum_{l=0}^1 \{-{M}({{B}^\T_0} x,a) \}^l {M}^{[1]}_{1-l}({{B}^\T_0} x,a) / {M}({{B}^\T_0} x,a),
\$
then we assume $\operatorname{det}(I_{M,B_0}) \neq 0$, where the information matrix $I_{M,B_0} = E\{M^{[1]}{({{B}^\T_0} X ,A)}^{\otimes2}\}$.
\end{assump}

\begin{assump} \label{assump3}
The bandwidth $h$ satisfy $hn^{1/\{2(\alpha-1)\}} \rightarrow 0$ and $n^{1/2}h^{d+3}/\log n\rightarrow 0$ for some $ \alpha \geq \lfloor (d+1)/2 \rfloor + 1$, where $\alpha \in \mathbbm{Z}^{+}$.
\end{assump}

\begin{thm}\label{thm:psudo}
Under Assumptions \ref{assump1}-\ref{assump3} and the model assumption given in  \eqref{eq:index}, we have that the estimator of the pseudo-direct learning $\text{vecl}(\widehat{B})$ is asymptotically normal,
\$
\sqrt{n}\big\{\text{vecl}(\widehat{B}) -  \text{vecl}({{B}_0})\big\}  \xrightarrow{}  \mathcal{N}\bigg(0,E\Big[\big\{\text{vecl}(I^{-1}_{M,{{B}_0}}S_{M,{{B}_0}})\big\}^{\otimes2}\Big] \bigg),
\$
where $\text{vecl}(\cdot)$ represents vectorization of the lower block of a matrix, and $S_{M,{{B}_0}} = \linebreak M^{[1]}({{B}^\T_0} X,A)\{R -{M}({{B}^\T_0} X,A)\}$ with $M^{[1]}({{B}^\T_0} X,A)$ and $I_{M,{{B}_0}}$ defined in Assumption \ref{assump2}.
\end{thm}

Theoretical results of the direct learning and direct-split learning approaches are significantly different from the pseudo-direct learning approach. It is interesting that standard results of semiparametric theory cannot be directly applied to obtain the asymptotic normality. This is mainly because of the unknown nonparametric function $f$ involved in the formulation. Some identifiability and smoothness conditions are required to develop the theoretical results. We state them after introducing the following definitions. Let $\text{vecl}({B}) \in \Theta$, where $\Theta$ is a subset of $\mathbb{R}^{(p-d)\times d}$. For any $\delta_1 >0 $, we define $\Theta_{\delta_{1}} = \{\text{vecl}({B}) \in \Theta: \|\text{vecl}({B})-\text{vecl}({B}_0) \| \leq \delta_{1}\}$. Further we let $\eta(x, B, f) = E\{R \mid B^\T X = B^\T x, A =f({B}^\T x) \}$ and let $\eta_0(x, B) = E\{R \mid B^\T X = B^\T x, A =f_{\text{opt}}(B^\T x) \} $.

\begin{assump}\label{assump4}
$B_0$ is a unique maximizer of $E\{\eta_0(X, {B})\}$, and $\text{vecl}({B}_0)$ is an interior point in $\Theta$ which is a compact subset of $\mathbb{R}^{(p-d)\times d}$.
\end{assump}

\begin{assump}\label{assump5}
Define $\mathcal{C}^{\alpha}_{M}(\cdot)$ as a class of functions on a bounded set with uniformly bounded partial derivatives up to order $ \lfloor \alpha \rfloor$.
In addition, the highest partial derivatives of such class of functions possess Lipschitz of order $\alpha- \lfloor \alpha \rfloor$. Assume for any $\text{vecl}(B) \in \Theta_{\delta_1}$, $\eta(x,B,f) \in \mathcal{C}^{\alpha}_{M}(\mathcal{X}_d)$, where $\mathcal{X}_d$ is a finite union of bounded and convex subsets of $\mathbb{R}^{d}$ with nonempty interior.
\end{assump}

Assumption \ref{assump4} is a standard assumption to ensure identifiability. Assumption \ref{assump5} imposes the smoothness condition for $\eta(x,B,f)$. We now present the consistency result for the direct learning approach implemented through Algorithm \ref{alg:direct}.

\begin{thm}\label{thm:direct:consist}
Let $\text{vecl}(\widehat{B})$ be the estimator of the direct learning approach implemented in Algorithm \ref{alg:direct}. Then under Assumption \ref{assump3}-\ref{assump5}, $\text{vecl}(\widehat{B})$ is consistent to $\text{vecl}({B_0})$, that is
\$
\big\|\text{vecl}(\widehat{B}) - \text{vecl}({B}_0)\big\| = o_{p}(1),
\$
where $\| \cdot \|$ is the is the $L_2$-norm.
\end{thm}

Analyzing the asymptotic distribution of the direct learning estimator $\text{vecl}({\widehat{B}})$ is slightly more involved. We needed to capture the effect of the nonparametric estimation of $\eta_0(x, {B})$ which includes two components: the conditional mean function $\eta(x, B, f)$, and the dose rule function $f(B^\T x)$ as one of the arguments in $\eta(x, B, f)$. This distinguishes us from the standard seimparametric estimation problems \citep{newey1994asymptotic,andrews1994asymptotics,ichimura2010characterization}, which usually only need to capture the effect of the conditional mean function but not the estimation of $f$. On the original non-splitting procedure, the samples used for estimating $f(B^\T x)$ is also for the kernel approximation of $\eta(x,B,f)$. Therefore, the covariance between $\widehat f(B^\T x)$ and $\widehat \eta(x,B,f)$ are non-zero, which will be involved in the reminder term in the asymptotic linear form of $\widehat \eta(x, B, f)$ and $\partial \widehat{\eta}(x, B,f)/\partial \text{vecl}(B)$. Without the fast convergence rate of the reminder terms, the asymptotic normality of $\text{vecl}({\widehat{B}})$ is hard to be guaranteed \citep{ichimura2010characterization}. To tackle this issue, we consider a cross-fitting strategy, also known as sample splitting \citep{bickel1982adaptive,powell1989semiparametric, 10.1111/ectj.12097,newey2018cross}. With the sample splitting procedure provided by Algorithm \ref{alg:split} in Supplementary Material, the variation of the estimation of $f(B^\T x)$ can be evaluated on the samples which are independent from the samples used for the kernel approximation of $\eta(x,B,f)$. This allows us to control the corresponding remainder term in the linear approximation, thus the asymptotic normality of $\text{vecl}({\widehat{B}})$ can be established. The following regularity assumptions are needed before we proceed to present the root-$n$ asymptotic normality result in Theorem \ref{thm:split:normal}. These conditions mainly concern the smoothness of $\eta_0(x, B)$ and its derivative.

\begin{assump}\label{assump6}
As a function of $B$, $\eta_0(x,B)$ is twice continuously differentiable on $\Theta_{\delta_1}$ with bounded derivatives.
\end{assump}

\begin{assump}\label{assump7}
For any $\text{vecl}(B) \in \Theta$,  $\partial \eta_0(x,B)/ \partial \text{vecl}(B)$ is twice continuously differentiable with respect to $x$.
\end{assump}

\begin{thm}\label{thm:split:normal}
Let
\$
V_0(X)&= E\bigg\{\frac{\partial^2\eta_0(X,B)}{\partial\text{vecl}(B) \partial\text{vecl}(B)^{\T} }\bigg\} \bigg |_{\text{vecl}(B) = \text{vecl}(B_0)}\\
\text{and} \quad \Gamma_0(X) &= \bigg[ \frac{\partial \eta_0(X,B)}{ \partial \text{vecl}(B)}- E\bigg\{\frac{\partial \eta_0(X,B)}{ \partial \text{vecl}(B)} \bigg\}\bigg]  \bigg|_{\text{vecl}(B) = \text{vecl}(B_0)}.
\$
Under the assumptions \ref{assump3}-\ref{assump7}, and we assume that $\Omega_0 = E\{\Gamma_0(X) \Gamma_0(X)^\T \}$ exists and $V_0$ is a positive definite matrix, then the estimator from the direct-split learning algorithm is asymptotically normal, that is
\$
\sqrt{n}\big\{\text{vecl}(\widehat{B}) - \text{vecl}({B}_0)\big\} \rightarrow \mathcal{N}\big(0,\Sigma \big),
\$
where $\Sigma = V^{-1}_0\Omega_0 V^{-1}_0$.
\end{thm}

To conclude this section, we present the convergence rate of $V_{\widehat f_{\text{opt}}}\big( \widehat {B} \big)$ to the optimal value function $V_{f_{\text{opt}}}({B}_0)$. A smoothness assumption is needed before we present Theorem \ref{thm:value}.

\begin{assump} \label{assump8}
For any $\text{vecl}(B) \in \Theta_{\delta_1}$, the conditional expectation $E(R \mid B^\T X = B^\T x, A=a)$ is Lipschitz continuous over $a$ with the Lipschitz constant independent of $a$.
\end{assump}

\begin{thm}\label{thm:value}
Under Assumptions \ref{assump3}-\ref{assump8}, the estimated value function based on estimators from the
direct-split learning algorithm converges to the optimal value function, that is
\$
V_{f_{\text{opt}}}({B}_0) - V_{\widehat f_{\text{opt}}}(\widehat { B}) = O_p\Bigg\{\bigg(\frac{\log n}{n}\bigg)^{\frac{\alpha}{2\alpha+d+1}}\Bigg\}.
\$
\end{thm}
\begin{rmk}
As $d \ll p$, Theorem \ref{thm:value} implies that the convergence rate of $V_{\widehat f_{\text{opt}}}(\widehat { B})$ is faster than the corresponding value function convergence rate $O_p\{n^{-1/(4+3p/\alpha)}\}$ in \citep{chen2016personalized}. When $d = p$, we can achieve the convergence rate $O_p\{(\log n/n)^{\alpha/(2\alpha+p+1)}\}$, which is still faster than the rate in \cite{chen2016personalized}.
\end{rmk}

\begin{rmk}
For the pseudo direct learning approach, the convergence rate of $V_{\widehat f_{\text{opt}}}\big( \widehat {B} \big)$ depends on the existing individualized dose-finding methods used in the second stage. For example, if we use the method of \cite{chen2016personalized}, the corresponding convergence rate should be faster since it enjoys the advantage of the proposed dimension reduction framework, that is $O_p\{n^{-1/(4+3d/\alpha)}\}$, where  $d \ll p$.
\end{rmk}

\section{Simulation Studies}\label{sec:sim}

We consider six different simulation settings. In the settings 1-4, data are generated from a randomized trial where the dose assignment $A \sim \text{Unif}[0,2]$, while in settings 5 and 6, observational data is considered. For each setting, we consider $p = 10$ or $20$. For all settings, we use $B = (\beta_1, \beta_2)$ where $\beta_1 = (1,0.5,0,0,-0.5, 0, \ldots, 0)^\T$ and $\beta_2 = (0.5,0,0.5,-0.5,1, 0, \ldots, 0)^\T$. The number of $0$ elements depends on the size of $p$. In setting 1, $X$ is generated independently from $\text{Unif}[-1,1]$. In settings 2, 3 and 5, $X$ is generated independently from a standard normal distribution. In setting 4, $X \sim N(0, \Sigma)$, where $\Sigma=(0.5^{|i-j|})_{ij}$. In setting 6, $X$ are independent normal with mean $2.5$ and variance 1. In setting 5, we let $A$ follow $\text{TruncNormal}(0.25 - 0.25|\beta_1^\T X| + \{0.75\beta_1^\T X + 0.75\}^{-1},1,2, 0)$, meaning that the propensity score is aligned with one of the dimensions of the treatment decision. Here, $\text{TruncNormal}(\mu,\sigma,u,l)$ is a truncated normal distribution with mean $\mu$, standard deviation $\sigma$, upper and lower bounds $u$ and $l$, respectively. In setting 6, $A$ follows $\text{Beta}(6.5-X_1 + 2X_4+X_7, 6.5)$, hence the treatment assignment is not in the dimension reduction subspace. The outcome $R$ is generated from a normal distribution with unit variance and a mean function $M(X, B, A)$. Both $M(X, B, A)$ and the optimal dose rule $f_{\text{opt}}$ are specified below. $f_{\text{opt}}$ is based on $d = 1$ in all settings except the setting 1.

\vspace{0.1cm}
\noindent Setting $1$: $f_{\text{opt}} = 0.6\mathbbm{1}(\beta_1^\T X > -0.6)\mathbbm{1}(\beta_2^\T X < 0.6) + 0.7\log(|\beta_1^\T X| + 0.5) + 0.5,$\\
\noindent $ M(X, B, A) = 6 + 0.3\log(|\beta_1^\T X| + 0.5) + \mathbbm{1}(\beta_2^\T X < 0.2) +2\mathbbm{1} (\beta_2^\T X > -0.7) - 16(f_{\text{opt}} - A)^2 $

\noindent Setting $2$: $f_{\text{opt}} = {3}\{{5(\beta_1^\T X)^2 + 2.5}\}^{-1} + \{(\beta_1^\T X)^4 + 1.3\}^{-1},$\\
\noindent $M(X, B, A) = -8 + 0.5|\beta_2^\T X| + 3.5\cos(\beta_2^\T X) + 15 \exp\big\{-(f_{\text{opt}}-A)^4\big\}.$

\noindent Setting $3$:  $f_{\text{opt}} = {0.7}({|\beta_2^\T X|/2 + 1})^{-1} + 1.5\log(|\beta_2^\T X|+1)-0.6,$ \\
\noindent $M(X, B, A) = -5 + 1.5\sin(\beta_1^\T X) + 3\cos(\beta_1^\T X) + 12 \exp\{-(f_{\text{opt}}-A)^2\}.$

\noindent Setting $4$:  $f_{\text{opt}} = 0.5\exp({-|\beta_1^\T X|}) + \sin(\beta_1^\T X) + 0.9,$ \\
\noindent $M(X, \beta_1, A) = 7 + 0.5(\beta_1^\T X)^2 + 0.5|\beta_1^\T X| + 4.5\cos(\beta_1^\T X) - 7|f_{\text{opt}} - A|.$

\noindent Setting $5$:  $f_{\text{opt}} = 0.5 + 0.5\cos(\beta_1^\T X) + \{1 + (\beta_1^\T X)^4\}^{-1},$ \\
\noindent $M(X, B, A) = -4 + \log\{\cos(\beta_2^\T X)  +1\} + 2.5\cos(\beta_2^\T X) + 13 \exp\{-(f_{\text{opt}}-A)^4\}.$

\noindent Setting $6$:  $f_{\text{opt}} = 0.25 + 0.125\{{\sin(\beta_1^\T X) + \cos(\beta_1^\T X)}\} + {0.25}\{{\exp(\beta_1^\T X) + 1}\}^{-1},$ \\
\noindent $M(X,B, A) = 8 + 0.5\sin(\beta_2^\T X) + 3 \mathbbm{1}(\beta_2^\T X<2.5) - 15|f_{\text{opt}} - A|.$

Each experiment is repeated $100$ times with a training sample size $n=400$ and testing sample size $3000$. We compare our methods with K-O-learning \citep{chen2016personalized}, Lasso \citep{tibshirani1996regression} and random forests \citep{liaw2002classification}. Since K-O learning requires positive $R$ values, and all other methods are invariance under intercept changes, we will force $R$ to be positive by subtracting its lowest value. Lasso and random forests are regression-based approaches that model the outcome $R$ first, then select the best dose level by maximizing the predicted outcome for a new subject. $\{X, A, X^2, X \!\cdot\! A, A^2\}$ is used as predictors for Lasso, and $\{X, A\}$ is used for random forests since the model automatically incorporates interactions. The random forest approach can be viewed as a continuous dose version of the virtual-twin model \citep{foster2011subgroup}. The code of K-O-learning is provided by \cite{chen2016personalized}. For Lasso, we used the R package ``glmnet'' \citep{Jerome2010}, and we implement the random forests approach using ``randomForest'' \citep{Andy2002} package. Tuning parameters are selected by $10$-fold cross-validation except $\phi_n$ in K-O-learning is fixed to be $0.15$, similar in \cite{chen2016personalized}.

Settings 5 and 6 are both observational studies. Our direct and direct-split learning approaches do not require to estimate the propensity score. For the pseudo direct learning and K-O-learning approaches, we use a boosting method \citep{zhu2015boosting} and Beta regression method \citep{zeileis2010beta} to estimate the propensity score. The propensity score model is misspecified in setting 5 while correctly specified in setting 6 with the Beta regression. We also evaluate pseudo-direct learning and K-O-learning approaches when ignoring the propensity score adjustment. To measure numerical performance of the estimated dose rule, the predicted value function $V_f(B)$ \citep{zhao2012estimating}, and the squared error of the estimated optimal dose, defined as $E\{ (\widehat f_{\text{opt}} - f_{\text{opt}})^2 \}$, are considered. These results are summarized in Table \ref{tab:1} and Table \ref{tab:2}.

We also investigate the accuracy of the estimated column space of the dimension reduction matrix $B$. We mainly compare the proposed methods with \cite{feng2013partial}, as it is one the few available approach that has the potential to detect iterations between $X$ and $A$. Furthermore, we analyze the robustness of the proposed approaches when $d$ is overspecified. In particular, we consider using $d = 2$ in setting 4. These results are presented and discussed in Supplementary Material.

\begin{table}[htp]
\footnotesize
\renewcommand{\arraystretch}{0.85}
    \caption{ Simulation results: mean (sd) predicted value function and mean (sd) squared dose distance }\center{
    \begin{tabular}{l c c c c }
& \multicolumn{2}{c}{predictor dimension $p=10$}&\multicolumn{2}{c}{predictor dimension $p=20$}\\
\noalign{\smallskip}
Method &   Predicted Value Function & Dose distance & Predicted Value Function & Dose distance \\
\noalign{\smallskip}
Setting 1\\
\noalign{\smallskip}
Direct      &           6.16 (\!\!    0.25    \!\!) &  0.07 (\!\!    0.02    \!\!)    &    5.45 (\!\!    0.44    \!\!) &    0.11    (\!\!    0.02    \!\!) \\
Direct-Split  &           5.71 (\!\!    0.33    \!\!) &  0.09 (\!\!    0.02    \!\!)    &    4.89 (\!\!    0.41    \!\!) &    0.14   (\!\!    0.03    \!\!) \\
Pseudo-Direct   &           6.52 (\!\!    0.14    \!\!) & 0.04     (\!\!    0.01    \!\!)    &    6.41 (\!\!    0.12    \!\!) &    0.05    (\!\!    0.01    \!\!) \\
K-O-learning          &           4.78 (\!\!    0.28    \!\!) & 0.16 (\!\!    0.02    \!\!)    &    4.06 (\!\!    0.25    \!\!) &    0.19    (\!\!    0.02    \!\!) \\
Random Forests      &           4.10 (\!\!    0.19    \!\!) & 0.18 (\!\!    0.01    \!\!)    &    3.68 (\!\!    0.40    \!\!) &    0.22    (\!\!    0.03    \!\!) \\
Lasso         &           3.05 (\!\!    3.17    \!\!) & 0.26     (\!\!    0.20 \!\!)    &    -5.83 (\!\!    1.96    \!\!) &    0.82    (\!\!    0.12    \!\!) \\
\noalign{\smallskip}
Setting 2\\
\noalign{\smallskip}
Direct     &           9.77 (\!\!    0.18    \!\!) &  0.08 (\!\!    0.03    \!\!)    &    9.41 (\!\!    0.31 \!\!) &    0.13    (\!\!    0.04    \!\!) \\
Direct-Split  &           9.60 (\!\!    0.40    \!\!) &  0.11 (\!\!    0.05    \!\!)    &    9.23 (\!\!    0.36    \!\!) &    0.15  (\!\!    0.04    \!\!) \\
Pseudo-Direct            &           9.64 (\!\!    0.21    \!\!) & 0.08 (\!\!    0.03    \!\!)    &    9.57 (\!\!    0.30    \!\!) &    0.10    (\!\!    0.05    \!\!) \\
K-O-learning          &           8.12 (\!\!    0.22    \!\!) & 0.28     (\!\!    0.03    \!\!)    &    7.27 (\!\!    0.18    \!\!) &    0.38    (\!\!    0.02    \!\!) \\
Random Forests      &           7.60 (\!\!    0.30    \!\!) & 0.35     (\!\!    0.03    \!\!)    &    7.12 (\!\!    0.36    \!\!) &    0.40    (\!\!    0.04    \!\!) \\
Lasso         &           2.78 (\!\!    1.34    \!\!) & 1.22     (\!\!    0.27    \!\!)    &    2.39 (\!\!    0.17    \!\!) &    1.30    (\!\!    0.04    \!\!) \\
\noalign{\smallskip}
Setting 3\\
\noalign{\smallskip}
Direct     &           7.69 (\!\!    0.31    \!\!) &  0.07 (\!\!    0.04    \!\!)    &    7.38 (\!\!    0.34    \!\!) &    0.10    (\!\!    0.04    \!\!) \\
Direct-Split  &           7.54 (\!\!    0.45    \!\!) &  0.09 (\!\!    0.05    \!\!)    &    7.22 (\!\!    0.42    \!\!) &    0.12   (\!\!    0.05   \!\!) \\
Pseudo-Direct            &           7.85 (\!\!    0.17    \!\!) & 0.06 (\!\!    0.03    \!\!)    &    7.76 (\!\!    0.26    \!\!) &    0.07    (\!\!    0.04    \!\!) \\
K-O-learning          &           6.94 (\!\!    0.13    \!\!) & 0.14 (\!\!    0.02    \!\!)    &    6.63 (\!\!    0.08 \!\!) &    0.18    (\!\!    0.01    \!\!) \\
Random Forests      &           6.45 (\!\!    0.10    \!\!) & 0.21 (\!\!    0.02    \!\!)    &    6.21 (\!\!    0.16    \!\!) &    0.24    (\!\!    0.02    \!\!) \\
Lasso         &           5.20 (\!\!    1.68    \!\!) & 0.42     (\!\!    0.30    \!\!)    &    3.15 (\!\!    0.26    \!\!) &    0.86    (\!\!    0.06    \!\!) \\
\noalign{\smallskip}
Setting 4\\
\noalign{\smallskip}
Direct     &           7.95 (\!\!    0.21    \!\!) &  0.18 (\!\!    0.04    \!\!)    &    7.69 (\!\!    0.18    \!\!) &    0.24    (\!\!    0.04    \!\!) \\
Direct-Split &           7.78 (\!\!     0.25    \!\!) &  0.26 (\!\!   0.06    \!\!)    &    7.56 (\!\!    0.21    \!\!) &    0.31  (\!\!    0.05   \!\!) \\
Pseudo-Direct           &           8.34 (\!\!    0.19    \!\!) & 0.14 (\!\!    0.02    \!\!)    &    8.16 (\!\!    0.15    \!\!) &    0.16    (\!\!    0.02    \!\!) \\
K-O-learning          &           6.85 (\!\!    0.18    \!\!) & 0.38     (\!\!    0.03    \!\!)    &    6.43 (\!\!    0.20    \!\!) &    0.42    (\!\!    0.04    \!\!) \\
Random Forests      &           7.30 (\!\!    0.20    \!\!) & 0.32 (\!\!    0.04    \!\!)    &    7.12 (\!\!    0.16    \!\!) &    0.33    (\!\!    0.03    \!\!) \\
Lasso         &           6.96 (\!\!    0.10    \!\!) & 0.36     (\!\!    0.02    \!\!)    &    6.81 (\!\!    0.11    \!\!) &    0.38    (\!\!    0.02    \!\!) \\
    \end{tabular}}\label{tab:1}
\end{table}

\begin{table}[htp]
\footnotesize
\renewcommand{\arraystretch}{0.85}
    \caption{ Simulation results for observational studies: mean (sd) predicted value function and mean (sd) squared dose distance }
    \begin{threeparttable}
    \makebox[{\textwidth}]{%
    \begin{tabular}{l c c c c }
& \multicolumn{2}{c}{predictor dimension $p=10$}&\multicolumn{2}{c}{predictor dimension $p=20$}\\
\noalign{\smallskip}
Method &   Predicted Value Function & Dose distance & Predicted Value Function & Dose distance \\
\noalign{\smallskip}
Setting 5\\
\noalign{\smallskip}
Direct  &           9.38 (\!\!    0.61  \!\!) & 0.16 (\!\!    0.09    \!\!)    &   8.83 (\!\!     0.82    \!\!) &    0.25    (\!\!   0.13    \!\!) \\
Direct-Split  &         9.05 (\!\!   0.84   \!\!) &  0.21 (\!\!    0.13    \!\!)    &   8.64 (\!\!    0.83    \!\!) &    0.27   (\!\!    0.13   \!\!) \\
Pseudo-Direct-Adjust       &      8.67 (\!\!    0.80   \!\!) & 0.30   (\!\!    0.19    \!\!)    &   7.89 (\!\!    1.19    \!\!) &    0.44    (\!\!    0.29    \!\!) \\
Pseudo-Direct          &           8.85 (\!\!    0.65    \!\!) & 0.23   (\!\!    0.11    \!\!)    &   8.65 (\!\!    0.95    \!\!) &     0.26   (\!\!    0.17    \!\!) \\
K-O-learning-Adjust         &           7.04 (\!\!    0.55    \!\!) & 0.49 (\!\!    0.09    \!\!)    &    6.18 (\!\!    0.53    \!\!) &    0.59   (\!\!    0.08  \!\!) \\
K-O-learning        &           7.67 (\!\!    0.56   \!\!) &   0.39 (\!\!    0.08   \!\!)    &     6.60 (\!\!    0.61   \!\!) &    0.53   (\!\!      0.08    \!\!) \\
Random Forests      &          6.98  (\!\!    0.54  \!\!) & 0.52   (\!\!    0.10    \!\!)    &   6.87 (\!\!    0.58    \!\!) &   0.55   (\!\!    0.11    \!\!) \\
Lasso         &            5.63 (\!\!    0.11    \!\!) &  0.81    (\!\!    0.02    \!\!)    &     5.55 (\!\!    0.27  \!\!) &    0.84    (\!\!    0.05    \!\!) \\
\noalign{\smallskip}
Setting 6\\
\noalign{\smallskip}
Direct  &           5.42 (\!\!    0.87  \!\!) & 0.06 (\!\!    0.03    \!\!)    &   4.98 (\!\!     0.98    \!\!) &     0.08    (\!\!   0.04    \!\!) \\
Direct-Split  &           5.35 (\!\!   0.94    \!\!) &  0.06 (\!\!    0.03    \!\!)    &  5.12 (\!\!    0.92    \!\!) &    0.07   (\!\!    0.03   \!\!) \\
Pseudo-Direct-Adjust       &     5.13 (\!\!    0.27  \!\!) & 0.06  (\!\!    0.01    \!\!)    &   4.84 (\!\!    0.26    \!\!) &    0.07    (\!\!    0.01    \!\!) \\
Pseudo-Direct          &           5.01 (\!\!    0.22    \!\!) & 0.06  (\!\!    0.01    \!\!)    &   4.65 (\!\!    0.19    \!\!) &     0.08   (\!\!    0.01    \!\!) \\
K-O-learning-Adjust         &           4.65 (\!\!    0.22    \!\!) & 0.08 (\!\!    0.01    \!\!)    &    4.46 (\!\!    0.17    \!\!) &    0.09  (\!\!    0.01  \!\!) \\
K-O-learning        &         4.50 (\!\!    0.23   \!\!) &   0.09 (\!\!   0.01   \!\!)    &     4.33 (\!\!    0.17   \!\!) &    0.10   (\!\!      0.01    \!\!) \\
Random Forests      &      4.21 (\!\!    0.59  \!\!) &  0.11   (\!\!    0.02    \!\!)    &   4.05 (\!\!    0.37    \!\!) &   0.11  (\!\!   0.01   \!\!) \\
Lasso         &           4.06 (\!\!    0.43    \!\!) &  0.11   (\!\!    0.02    \!\!)    &     3.94 (\!\!    0.05  \!\!) &    0.12   (\!\!    0.01   \!\!) \\
\end{tabular}}\label{tab:2}
\bigskip
\vspace{0.1cm}
{\scriptsize Note: ``Pseudo-Direct-Adjust'' and ``K-O-learning-Adjust'' are the methods using the estimated $p(A|X)$.}
\end{threeparttable}
\end{table}

\setlength{\textfloatsep}{5pt}

As the results of Table \ref{tab:1}-\ref{tab:2}, the proposed three approaches achieve significantly better performance compared to existing methods. The performance of direct and direct-split learning is similar. When the outcome function depends on two directions, and the dose rule $f_{\text{opt}}$ depends on only one direction, the two direct learning approaches perform similar or better than the pseudo-direct learning. This is mainly because the direct learning approaches only estimate the subspace of $f_{\text{opt}}$. Hence, they are more efficient than pseudo-direct learning, which estimates all directions related to the outcome. Among competing methods, the K-O-learning has the best performance, followed by random forests and Lasso. However, K-O-learning is relatively sensitive to the dimension of covariates, suffering more under $p=20$ compared with $p=10$. For random forests, another disadvantage is the computational cost. In the settings with $d=2$, the random forest costs several minutes to estimate the optimal dose level in a single experiment, while the proposed approaches require less than $30$ seconds for the same experiment. The Lasso method fails to estimate, as expected, the optimal rule in the non-linear settings. We also observe that the performance of K-O-learning is affected by the propensity score adjustment. In setting 5 where the propensity score model is misspecified, the biased estimation of the propensity score significantly damages the performance.  The direct learning, including direct-split learning, outperforms K-O-learning by circumventing the bias from the model misspecification. In setting 6 where the propensity score is correctly specified, K-O-learning-Adjustment has an improved performance comparing to K-O-learning. This implies that the correct specification and accurate estimation of the propensity score play an important role in the K-O-learning approach. However, even if the propensity score is correctly specified, the direct and direct-split learning can still achieve better performance; besides, they also avoid additional computation cost because they exclude the procedure of estimating the propensity score.

\section{Data Analysis}\label{sec:data}
Warfarin, commonly known as a ``blood thinner'', is one of the most broadly used oral anticoagulant agents to treat blood clots and prevent forming new harmful blood clots to decrease the risk of heart attack or stroke. The appropriate dose level of warfarin significantly affects the treatment effects. Hence, we apply our proposed methods to the dataset provided by \cite{international2009estimation} to estimate the optimal dose level. The International Warfarin Pharmacogenetics Consortium provided one of the most comprehensive and public datasets of clinical and pharmacogenetic covariates. We acquire both the pharmacogenetic and clinical data, including height, weight, age, race, phenytoin, carbamazepine, amiodarone, VKORC1 genotype, and CYP2C9 genotype. To measure the reward, we consider $\mathcal{R} = -|2.5-\text{International Normalized Ratio}|$, where the international normalized ratio is the primary outcome to measure the safety and efficiency of the dose level of warfarin. For patients prescribed warfarin, the target INR is around $2.5$. After excluding missing observations, we obtain the data with $2344$ patients.

It has been shown in \cite{international2009estimation} that the data were collected from observational studies instead of a randomized dose assignment. Hence, when necessary, such as the pseudo-direct learning and K-O-learning, we adjust for the propensity score $P(A \mid X)$ using the same approach described in \cite{zhu2015boosting}. Determining the structural dimension is a practically important task. Several approaches are possible. For example, we may treat the number of dimensions as a tuning parameter and select it through cross-validation. Another possible approach is to perform model selection using a BIC type of criterion in \citep{zhu2006sliced,feng2013partial} for determining the structure dimension. Besides, we can also follow a modified information criterion method developed in \cite{ma2015validated}. The details are given in Supplementary Material. After applying this method, we found that $d=1$ achieves the best fitting performance for all proposed approaches. Hence, for all subsequent analyses, we set $d=1$.

To make a comparison among the methods, we randomly split the dataset $100$ times to obtain the training dataset including $800$ patients and testing dataset consisting of the rest $1544$ patients. However, in practice, the true dose level is still unknown on the testing set. Hence, there does not exist a direct measure to evaluate the performance. To address this issue, we calculate an expected reward for each subject on the testing dataset. To be specific, suppose the predicted optimal doses are $a_1, \ldots, a_n$, respectively for subjects on the testing dataset, we then calculate an estimated value function $\widehat{\mathcal{R}}$ which is only based on the testing dataset,
\$
\widehat{\mathcal{R}}_{\text{test}} = \frac{1}{\text{ntest}}\sum\limits_{j=1}^\text{ntest}\frac{ \sum_{i=1}^\text{ntest} R_i K_{v}(X_j - X_i, a_j - A_i)}{\sum_{i=1}^\text{ntest} K_{v}(X_j -  X_i, a_j - A_i) }.
\$
\begin{figure}[h]
\centering
\includegraphics[width=15cm,height=8cm]{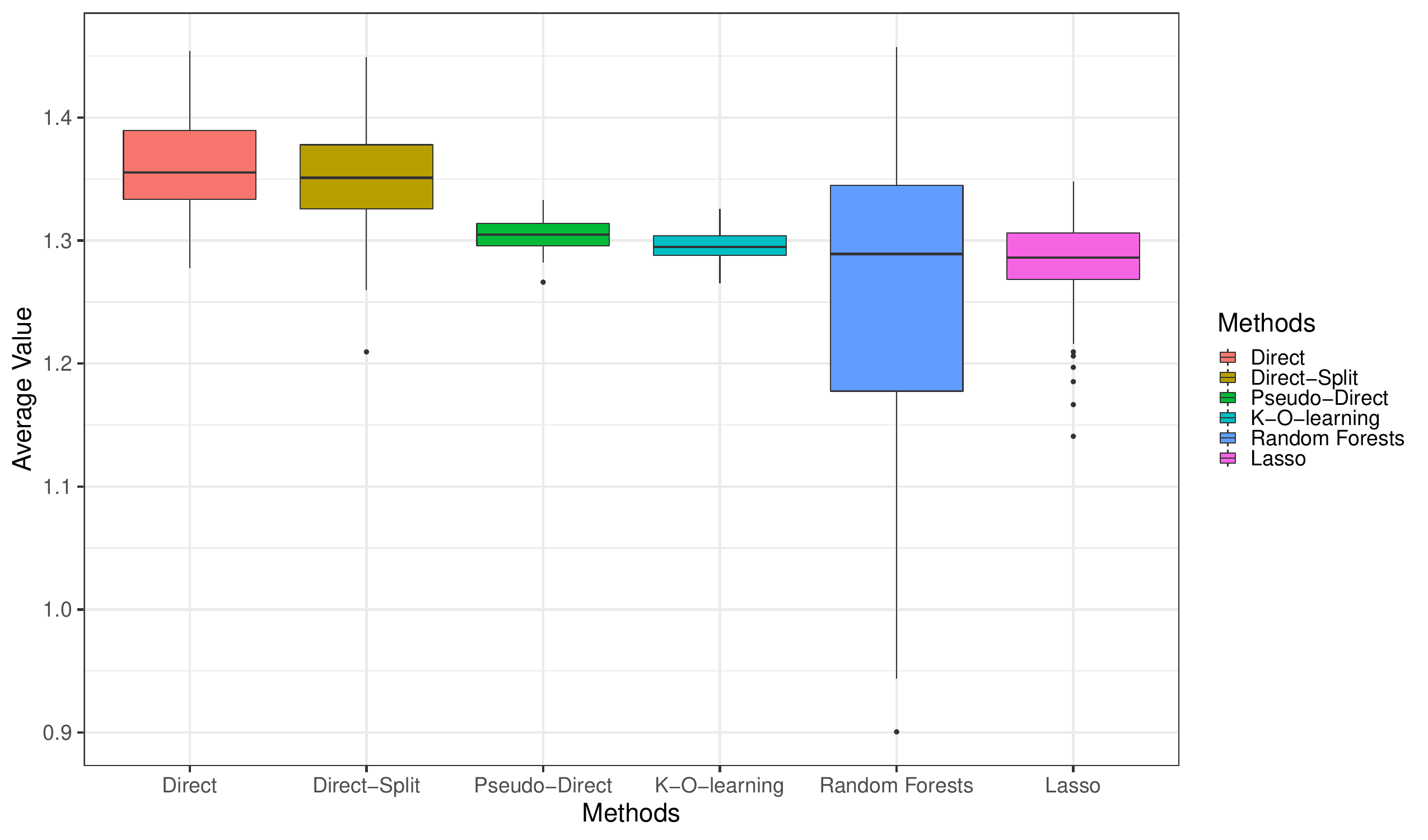}
\caption{{Boxplot of predicted value function $\widehat{\mathcal{R}}_{\text{test}}$}}\label{Fig:realdataplot}
\end{figure}

We report the average values and standard errors of the estimated value function on the testing dataset over repeating $100$ experiments in Figure \ref{Fig:realdataplot}. The results suggest that direct and direct-split learning approaches perform best, while the pseudo-direct learning approach slightly outperforms other methods. Most of the observed warfarin dose levels in the dataset are not far away from the optimal dose level because most observations have an observed INR close to $2.5$. Hence, the improvement we observe is minor.

\section{Discussion}\label{sec:diss}

When the number of covariates is extremely large, most sufficient dimension reduction methods may fail to estimate the subspace spanned by $B$ accurately. Hence, some additional modifications are probably needed \citep{wang2013sparse, wang2018estimating}. It would be interesting to extend the proposed dimension reduction framework to the right-censored outcomes \citep{zhao2014doubly,zhu2017greedy} and construct a corresponding direct learning approach to estimate the optimal dose rule. To estimate the covariance of the direct-split learning estimator in Theorem \ref{thm:split:normal}, we may use a sample estimator of $\widehat \Sigma = \widehat{V}_{n}^{-1} \widehat{\Omega}_{n} \widehat{V}_{n}^{-1}$. However, this requires to compute a second-order partial derivative in the term $\widehat{V}_{n}$. This can lead to unsatisfactory performance when the sample size is small. Alternatively, we can use a bootstrap estimator, and the results of $90\%$ confidence interval are provided in Supplementary Material. We observe a mild under-coverage at around $5\%$. Since the proposed objective function is non-convex, the result is satisfactory. However, additional work may still be required for further investigation to reduce the under-coverage.

For many complex diseases, a dynamic treatment regime that involves a sequence of decision rules is often needed. For example, \cite{rich2014simulating} proposed an adaptive strategy for the multiple-stage personalized dose-finding; \cite{zhao2015new} estimated the optimal dynamic treatment regime by converting the estimation problem to a single classification problem. Estimating the propensity score is again a challenging task. The proposed method may shed light under such settings.

\section{Acknowledgement}\label{sec:acknow}
The authors thank the editor, associate editor and two reviewers for their valuable comments. R. Zhu was partially supported by the National Center for Supercomputing Applications Fellowship and the University of Illinois at Urbana-Champaign research board grant RB19046. D. Zeng was partially supported by the U.S. National Institutes of Health grant GM124104.

\section{Supplementary Material}\label{sec:supp}
Supplementary material available at \textit{Biometrika} online includes the details of the warm start procedure, the direct-split algorithm, the determination of the structure dimension used in the real data analysis, the additional numerical results, the proofs of Theorems \ref{thm:psudo}-\ref{thm:value} and lemmas.

\section{Addition Results}\label{sec:results}

\subsection{Warm start of $w$ in the direct learning algorithm}

We use the approach motivated by the empirical location of the maximum procedures \citep{muller1985kernel}. The idea is as follows. For current iterating step $t$, we first take the current covariates ${B^{(t)}}^\T X$ to solve the following extrema problem for each subject $i = 1, \ldots, n$:
\#
\widetilde A_i = \operatorname*{argmax}_{\{a\}} \, \widehat R_i\big(a, B^{(t)}\big), \label{extrema}
\#
where a kernel approximated reward $\widehat R_i(a, B^{(t)})$ is defined as
\#
\widehat R_i(a, {B^{(t)}}) = \frac{ \sum_{j=1}^n R_j K_{v}\big\{\big({X^\T_j}  B^{(t)},A_j\big)^\T  -\big({X^\T_i}  B^{(t)},a\big)^\T \big\}}{\sum_{j=1}^n K_{v}\big\{\big({X^\T_j}  B^{(t)},A_j\big)^\T  -\big({X^\T_i}  B^{(t)},a\big)^\T \big\}}.
\label{eq:pseudo}
\#
Note that without the regularity term in \eqref{eq:sampleval}, the new update of $f({B^{(t)}}^\T X_i) = \widetilde A_i$ already maximizes the expected reward. By imposing the regularity term, we can solve \eqref{eq:foptim} in a class of smooth functions. This can be implemented by a kernel ridge regression approach:
\#
&~ \operatorname*{maximize}_{f \in {\cal H}} \,\, \sum^n_{i=1} \big\{\widetilde A_i - f({B^{(t)}}^\T X_i) \big\}^2 - \lambda_n \lVert f \rVert^2_{\cal H} \nonumber \\
=&~ \operatorname*{maximize}_{ w} \,\,  \sum^n_{i=1} \Big\{\widetilde A_i - \sum_j w_j K({B^{(t)}}^\T X_i, {B^{(t)}}^\T X_j) \Big\}^2 - \lambda_n \lVert  w \rVert^2,
\label{eq:ridge}
\#
where $K(\cdot, \cdot)$ is a $d$-dimensional kernel function, $f$ is from a Reproducing Kernel Hilbert Space $\cal H$ and $\|\cdot\|_{\cal H}$ is the corresponding norm. The analytic solution is given by
\#
f^{(t)}(\cdot) =& \sum^n_{j=1} w_j K(\cdot, {B^{(t)}}^\T X_j) = {\widetilde{A}}^\T ( K_{n \times n} + \lambda_n I)^{-1} \kappa(\cdot, {B^{(t)}}^\T {X}), \label{eq:fupdate}
\#
where $K_{n \times n}$ is an $n \times n$ kernel matrix with $K({B^{(t)}}^\T X_i, {B^{(t)}}^\T X_j)$ as the $(i, j)$'th entry, and $\kappa(\cdot, {B^{(t)}}^\T {X})$ is an $n \times 1$ vector with $K(\cdot, {B^{(t)}}^\T X_i)$ as the $i$'th element. Hence, the solution of ${w}$ is
\#
{w} =& (K_{n \times n} + \lambda_n I)^{-1} \widetilde{A}. \label{eq:w}
\#
The tuning parameter $\lambda_n$ can be chosen adaptively using the generalized cross-validation \citep{golub1979generalized}.

From a theoretical point of view, $\widetilde A_i$ should be obtained by maximizing over all possible values of dose, which may require a one-dimensional gradient descent of the function $\widehat R_i\big(\cdot, B^{(t)}\big)$. However, such an approach may not be necessary and is computationally inefficient. Instead, we can simply take a set of grid points on the dose range and optimize on the grid. This is mainly because that our primarily interest is the functional form of $f^{(t)}(\cdot)$ rather than the individual best dose value $\widetilde A_i$. Following the idea of \cite{muller1985kernel}, we can choose a set of $q = O(n^{1/2})$ grid points such that $\min{\{A_i\}} = a_1 < \cdots < a_q = \max{\{A_i\}}$, and solve for the maxima $\widetilde{A}_i$ on this grid as an approximation of the true maxima in \eqref{extrema}. This may result in an extra $O(n^{-1/2})$ approximation error, however, it should not affect the theoretical results since the rate of the approximation error is faster than the rate of a kernel ridge regression \citep{dicker2017kernel}. In practice, we adopt a uniform partitioning which fixes the difference $a_{j+1} - a_j$ as a constant for $j=1,2,...,q-1$. Hence, we obtain $\widetilde A_i$ as
\#
\widetilde A_i  = \operatorname*{argmax}_{a \in \{a_1, \ldots, a_q\} } \widehat R_i(a, {B^{(t)}}),
\label{eq:discrete}
\#
then the solution ${w}$ can be obtained using \eqref{eq:discrete} instead of \eqref{extrema}.

As we mentioned above, another way of solving problem \eqref{extrema} is to use a one-dimensional optimization, which may yield a more accurate result for each subject. It should be noted that accuracy for each subject is not a severe issue because the optimal dose rule is jointly modeled by all subjects in \eqref{eq:fupdate}. It can be computationally intensive due to the non-convexity of the objective function, and there is no significant difference in terms of the performance.

\subsection{Direct-split learning algorithm}

The following is a modification of the direct learning algorithm proposed in Algorithm \ref{alg:direct} provided in the main text. The major modification is to split the data into two separate sets ${\cal D}_1$ and ${\cal D}_2$, according to the description provided in Section \ref{sec:alg:split}.

\setcounter{algorithm}{1}

\begin{algorithm}\label{alg:split}
Direct-Split learning algorithm
\begin{tabbing}
\qquad \enspace {Initialize}: ${\cal D} = \{X_i, A_i, R_i\}_{i=1}^n$, $B^{(0)}$ such that ${B^{(0)}}^\T B^{(0)} = {I}$, $\varepsilon \leftarrow 10^{-8}$. \\
\qquad \enspace {Splitting}: Split ${\cal D}$ to ${\cal D}_1$ and ${\cal D}_2$ randomly.\\
\qquad \enspace {For} $t = 1$ to $t =$ max.iter:\\
\qquad \qquad Fixing $B^{(t)}$, calculate $R_i(a, B^{(t)})$ and $\widetilde{A}_i$ in equations \eqref{eq:pseudo} and \eqref{eq:discrete} using ${\cal D}_1$. \\
\qquad \qquad Update ${w}$ in $f^{(t+1)}(\cdot)$ based on equation \eqref{eq:fupdate} using ${\cal D}_1$.\\
\qquad \qquad Fixing $f^{(t+1)}(\cdot)$, numerically approximate $G$ and $Q$ based on equation \eqref{eq:grad} using ${\cal D}_2$.\\
\qquad \qquad Use a line search algorithm to find the best step size $\tau$ in equation \eqref{eq:update}.\\
\qquad \qquad Update $B^{(t+1)} \leftarrow {\big({I} + \frac{\tau}{2}Q\big)}^{-1}\big({I} - \frac{\tau}{2}Q\big)B^{(t)}$ using ${\cal D}_2$.\\
\qquad \qquad Stop if $ \lVert G \rVert \leq \varepsilon$.\\
\qquad \enspace {Return}: $\widehat B = B^{(t+1)}$ and $\widehat f_{\text{opt}} = f^{(t+1)}$.
\end{tabbing}
\end{algorithm}

\subsection{Asymptotic variance estimation}
In Theorem \ref{thm:split:normal}, we show that the estimator of direct-split learning $\text{vecl}({\widehat{B}})$ is asymptotic normal. We use the bootstrapping method to estimate its asymptotic covariance $\Sigma$. To evaluate performance of the covariance estimation, we calculate the $90\%$ confidence interval of $\text{vecl}(\widehat{B})$ in the simulation setting 2, where the true parameter for dose rule is $B = (1,0.5,0,0,-0.5, 0, 0,0,0, 0)^\T$. As we mention in Section \ref{sec:theory}, we consider an upper-block diagonal version of the parameter matrix $B$ following the idea in \citep{ma2013efficient}. Then the parametrized version of $B$ is $\text{vecl}(B) = \beta =   (0.5,0,0,-0.5, 0, 0,0,0, 0)^\T$. The results are summarized in Figure \ref{Fig:ci_cover} and Table \ref{tab:4}.

\setcounter{figure}{1}

\begin{figure}[h]
\centering
\includegraphics[width=14.5cm,height=9.5cm]{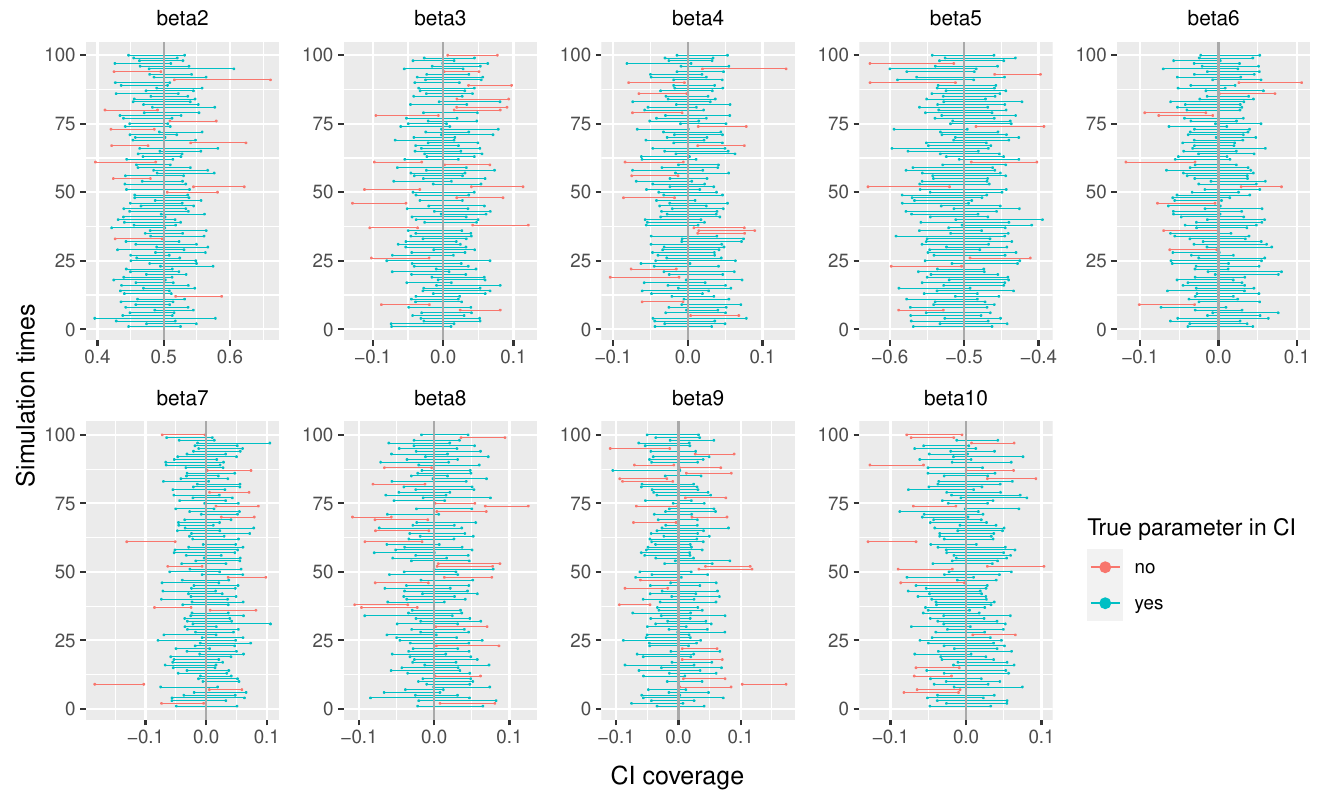}
\caption{{90\% bootstrap confidence interval estimation over 100 times simulations under simulation setting 4 with $p=10$ }}\label{Fig:ci_cover}
\end{figure}

\setcounter{table}{2}

\begin{table}[h]
\small
\renewcommand{\arraystretch}{0.85}
    \caption{{90\% bootstrap confidence interval covering times over 100 times simulations }}\center{
    \begin{tabular}{l c c c c c c c c c c }
Elements in $B$ & $B_{1}$ &  $B_{2}$ & $B_{3}$ & $B_{4}$ &  $B_{5}$ & $B_{6}$  & $B_{7}$ &  $B_{8}$ & $B_{9}$ & $B_{10}$\\
\noalign{\smallskip}
Parameter value &  1 &  0.5 & 0 & 0 & -0.5 & 0 & 0 & 0 & 0 & 0\\
\noalign{\smallskip}
Coverage \% &  - &  87 &  82 & 84 &91 & 90 & 87 & 80 & 79& 84\\
    \end{tabular}}\label{tab:4}
    \end{table}

\subsection{Additional simulation results}

This section contains additional simulation results of investigating the accuracy of the estimated column space of the dimension reduction matrix $B$. Performance is measured by three criteria: the Frobenius norm distance between the projection matrix $P_{B} = B(B^\T B)^{-1} B^\T$ and corresponding estimator's version $P_{\widehat B}$; the trace correlation $\textnormal{tr}\big(P_{B} P_{\widehat B}\big)/d$, where $d$ is the structural dimension; and the canonical correlation between $B^\T X$ and ${\widehat B}^\T X$. For the competing method, we implement partial-SAVE in \citep{feng2013partial}. The results are summarized in Table \ref{tab:3}. It should be noted that for settings 2 to 6, the direct learning approaches are compared against one true direction of the dose rule, while the pseudo-direct approach and partial-SAVE are compared with two true directions of the outcome.

\begin{table}[h]
\small
\renewcommand{\arraystretch}{0.85}
    \caption{{Simulation results: comparison for estimating the dimension reduction space}}\center{
    \begin{tabular}{l c c c c c c}
 & \multicolumn{3}{c}{predictor dimension $p=10$}&\multicolumn{3}{c}{predictor dimension $p=20$}\\
Method &   Frobenius & Trace & Canonical & Frobenius & Trace & Canonical \\
\noalign{\smallskip}
Setting 1\\
\noalign{\smallskip}
Direct     &    0.39    (\!\!    0.13    \!\!)    &    0.95    (\!\!    0.02    \!\!)    &    0.98    (\!\!    0.02    \!\!)    &    0.74    (\!\!    0.21    \!\!)    &    0.85    (\!\!    0.13    \!\!)    &    0.92    (\!\!    0.07    \!\!)    \\
Direct-Split    &    0.63   (\!\!    0.15    \!\!)    &    0.89    (\!\!    0.06    \!\!)    &    0.95    (\!\!    0.04    \!\!)    &    0.98    (\!\!    0.17    \!\!)    &    0.75    (\!\!    0.10    \!\!)    &    0.86   (\!\!    0.06    \!\!)    \\
Pseudo-Direct         &    0.19    (\!\!    0.07    \!\!)    &    0.99    (\!\!    0.02    \!\!)    &    0.99    (\!\!    0.01    \!\!)    &    0.29    (\!\!    0.08    \!\!)    &    0.98    (\!\!    0.02    \!\!)    &    0.99    (\!\!    0.02    \!\!)    \\
partial-SAVE &    0.62    (\!\!    0.10    \!\!)    &    0.94    (\!\!    0.03    \!\!)    &    0.99    (\!\!    0.02    \!\!)    &    0.97    (\!\!    0.15    \!\!)    &    0.84    (\!\!    0.08    \!\!)    &    0.87    (\!\!    0.05    \!\!)\\
\noalign{\smallskip}
Setting 2\\
\noalign{\smallskip}
Direct    &    0.21    (\!\!    0.08    \!\!)    &    0.97    (\!\!    0.01    \!\!)    &    0.98    (\!\!    0.01    \!\!)    &    0.36    (\!\!    0.09    \!\!)    &    0.93    (\!\!    0.03    \!\!)    &    0.97    (\!\!    0.01    \!\!)    \\
Direct-Split    &    0.27    (\!\!    0.12    \!\!)    &    0.95    (\!\!    0.07   \!\!)    &    0.97    (\!\!    0.05    \!\!)    &    0.40    (\!\!    0.08    \!\!)    &    0.91    (\!\!    0.04    \!\!)    &    0.96    (\!\!    0.02    \!\!)    \\
Pseudo-Direct            &    0.33    (\!\!    0.09    \!\!)    &    0.97    (\!\!    0.02    \!\!)    &    0.98    (\!\!    0.01    \!\!)    &    0.71    (\!\!    0.29    \!\!)    &    0.85    (\!\!    0.13    \!\!)    &    0.89    (\!\!    0.12    \!\!)    \\
partial-SAVE    &    0.63    (\!\!    0.20    \!\!)    &    0.89    (\!\!    0.08    \!\!)    &    0.94    (\!\!    0.05    \!\!)    &    1.04    (\!\!    0.25    \!\!)    &    0.71    (\!\!    0.13    \!\!)    &    0.81    (\!\!    0.13    \!\!)    \\
\noalign{\smallskip}
Setting 3\\
\noalign{\smallskip}
Direct    &    0.37    (\!\!    0.19    \!\!)    &    0.90    (\!\!    0.14    \!\!)    &    0.94    (\!\!    0.06    \!\!)    &    0.57 (\!\!    0.21    \!\!)    &    0.81    (\!\!    0.14    \!\!)    &    0.89    (\!\!    0.11    \!\!)    \\
Direct-Split    &     0.45    (\!\!    0.26    \!\!)    &    0.86   (\!\!    0.20    \!\!)    &    0.92    (\!\!    0.16    \!\!)    &    0.61    (\!\!    0.24    \!\!)    &    0.78    (\!\!    0.20    \!\!)    &    0.87    (\!\!    0.16    \!\!)    \\
Pseudo-Direct           &    0.13    (\!\!    0.03    \!\!)    &    0.99    (\!\!    0.01    \!\!)    &    0.99    (\!\!    0.01    \!\!)    &    0.21    (\!\!    0.04    \!\!)    &    0.98    (\!\!    0.01    \!\!)    &    0.99    (\!\!    0.01    \!\!)    \\
partial-SAVE    &    0.59    (\!\!    0.20    \!\!)    &    0.86    (\!\!    0.07    \!\!)    &    0.89    (\!\!    0.04    \!\!)    &    0.85    (\!\!    0.15    \!\!)    &    0.82    (\!\!    0.07    \!\!)    &    0.90    (\!\!    0.05    \!\!)    \\
\noalign{\smallskip}
Setting 4\\
\noalign{\smallskip}
Direct    &    0.59    (\!\!    0.10    \!\!)    &    0.95    (\!\!    0.04    \!\!)    &    0.96    (\!\!    0.02    \!\!)    &    0.72    (\!\!    0.10    \!\!)    &    0.75    (\!\!    0.09    \!\!)    &    0.83    (\!\!    0.04    \!\!)    \\
Direct-Split    &    0.71    (\!\!    0.16    \!\!)    &    0.73    (\!\!    0.13    \!\!)    &    0.86    (\!\!    0.08    \!\!)    &    0.87    (\!\!    0.16    \!\!)    &    0.61    (\!\!    0.16    \!\!)    &    0.82    (\!\!    0.09    \!\!)    \\
Pseudo-Direct            &    0.09    (\!\!    0.02    \!\!)    &    1.00    (\!\!    0.00    \!\!)    &    1.00    (\!\!    0.00    \!\!)    &    0.24    (\!\!    0.07    \!\!)    &    0.97    (\!\!    0.02    \!\!)    &    0.99    (\!\!    0.01    \!\!)    \\
partial-SAVE    &    0.73    (\!\!    0.10    \!\!)    &    0.76    (\!\!    0.07    \!\!)    &    0.87    (\!\!    0.02    \!\!)    &    0.93    (\!\!    0.17    \!\!)    &    0.56    (\!\!    0.17    \!\!)    &    0.79    (\!\!    0.08    \!\!)    \\
\noalign{\smallskip}
Setting 5\\
\noalign{\smallskip}
Direct    &     0.35    (\!\!    0.13    \!\!)    &    0.93    (\!\!    0.07    \!\!)    &     0.96    (\!\!   0.04    \!\!)    &     0.50 (\!\!  0.15    \!\!)    &    0.87  (\!\!    0.09   \!\!)    &    0.93    (\!\!    0.06  \!\!)    \\
Direct-Split    &    0.38    (\!\!    0.17    \!\!)    &    0.91    (\!\!    0.10    \!\!)    &    0.95    (\!\!    0.06   \!\!)    &    0.53    (\!\!    0.18    \!\!)    &    0.84   (\!\!    0.13    \!\!)    &    0.92    (\!\!    0.10    \!\!)    \\
Pseudo-Direct           &    0.21    (\!\!    0.06    \!\!)    &    0.99    (\!\!    0.01    \!\!)    &    0.99    (\!\!    0.00    \!\!)    &  0.46  (\!\!    0.31    \!\!)    &   0.92    (\!\!    0.13    \!\!)    & 0.95   (\!\! 0.11    \!\!)    \\
partial-SAVE    &   0.59  (\!\!    0.16    \!\!)    &    0.91    (\!\!    0.06    \!\!)    &    0.95    (\!\!    0.03    \!\!)    &     0.96   (\!\!    0.22    \!\!)    &  0.76    (\!\!  0.12    \!\!)    &    0.86   (\!\!    0.11   \!\!)    \\
\noalign{\smallskip}
Setting 6\\
\noalign{\smallskip}
Direct    &    0.55    (\!\!    0.23   \!\!)    &    0.82    (\!\!   0.16   \!\!)    &    0.95   (\!\!   0.13   \!\!)    &    0.83 (\!\!   0.25   \!\!)    &    0.62    (\!\!   0.23    \!\!)    &    0.84    (\!\!    0.24   \!\!)    \\
Direct-Split    &    0.55    (\!\!    0.21   \!\!)    &    0.83    (\!\!    0.15    \!\!)    &    0.95   (\!\!    0.12   \!\!)    &    0.82    (\!\!    0.27    \!\!)    &    0.63    (\!\!    0.25    \!\!)    &    0.83    (\!\!    0.25    \!\!)    \\
Pseudo-Direct           &    0.34    (\!\!    0.35    \!\!)    &    0.94    (\!\!    0.13    \!\!)    &    0.97    (\!\!    0.09    \!\!)    &   0.84    (\!\!    0.47    \!\!)    &    0.77   (\!\!     0.20   \!\!)    &    0.84    (\!\!   0.16   \!\!)    \\
partial-SAVE    &     0.94   (\!\!    0.26    \!\!)    &    0.76   (\!\!    0.13    \!\!)    &    0.86    (\!\!    0.11   \!\!)    &    1.28    (\!\!    0.19    \!\!)    &    0.58    (\!\!    0.12    \!\!)    &    0.77    (\!\!   0.13    \!\!)
    \end{tabular}}\label{tab:3}
    \end{table}

Overall, the proposed methods achieve better performance than partial-SAVE in \citep{feng2013partial}. In particular, the pseudo direct learning method achieves the best overall performance except for setting 2. This is probably due to the design of the mean reward function $M(X, B, A)$. When the mean reward function $M(X, B, A)$ has a dimension reduction structure, meaning that $M(X, B, A)$ is a function of $B^\T X$, the pseudo direct learning method should be the optimal one since it estimates the whole subspace. On the other hand, the direct learning approaches are under the risk of a greater bias when trying to estimate only one direction.

\subsection{Misspecification of the structure dimension}

In this example, we investigate the performance (robustness) of proposed methods under a wrong choice of $d$. Again, we consider setting 4, which is a one-dimensional setting. In the model specification, we force $d=2$ for all methods, hence an overfitting is likely to occur. The predicted value and dose distance are both slightly worse than the $d=1$ case presented in the main text. However, the difference is very minor. It should be noted that an under-specification that $d=0$ is likely to be more harmful to the performance. Hence, the selection of the structure dimension is an importance issue that worth further investigation. The method proposed in the next section, and the discussion section, can be considered in practice.

\begin{table}[h!]
\footnotesize
\renewcommand{\arraystretch}{0.85}
    \caption{Simulation results: misspecification of structure dimension (d=2) in Setting 4 }\center{
    \begin{tabular}{l c c c c }
& \multicolumn{2}{c}{predictor dimension $p=10$}&\multicolumn{2}{c}{predictor dimension $p=20$}\\
\noalign{\smallskip}
Method &   Predicted Value Function & Dose distance & Predicted Value Function & Dose distance \\
\noalign{\smallskip}
Direct     &       7.73 (\!\!    0.21    \!\!) &  0.27 (\!\!    0.05    \!\!)    &    7.49 (\!\!    0.19    \!\!) &    0.32   (\!\!    0.05    \!\!) \\
Direct-Split &           7.51 (\!\!     0.25    \!\!) &  0.35 (\!\!   0.10    \!\!)    &    7.29 (\!\!    0.20   \!\!) &    0.39  (\!\!    0.08   \!\!) \\
Pseudo-Direct           &           8.03 (\!\!    0.17    \!\!) & 0.17 (\!\!    0.02    \!\!)    &    8.00 (\!\!    0.19    \!\!) &    0.19    (\!\!    0.03    \!\!) \\
    \end{tabular}}\label{tab:misspecified}
\end{table}

\subsection{Structure dimension determination}

In the simulation study, we assume that the true structure dimension $d_0$ is known. The proposed methods can be modified to estimate $d_0$. We should note that for the pseudo-direct learning method, since the model in \eqref{eq:index} can be regarded as a set of nested semiparametric models indexed by $d$, the determination of true structural dimension $d_0$ naturally becomes a model selection problem. Therefore, we can follow a similar idea of the BIC-type of criterion \citep{zhu2006sliced,feng2013partial} to estimate the true structure dimension. Additionally, we may treat the structure dimension as a tuning parameter and select it through cross-validation. For the direct and direct-split learning approach, a modified information criterion method \citep{ma2015validated} could be applied. Note that we may modify the value function by incorporate an argument $d$, denoted as $V_{n,f}(B, d)$. And we further denote the partial derivative of $V_{n,f}(B,d)$ with respect to $\text{vecl}({B})$ as $V_{n,f}^{(1)}(B,d)$. It is reasonable to consider the framework that
\$
\widehat{d} &= \operatorname*{argmin}_{d} \Big\|V_{n,\widehat{f}_{\text{opt}}}^{(1)}(\widehat{B},d)\Big \|^2 + pd \log(n) \\
& \doteq \operatorname*{argmin}_{d}  \text{IC}(d),
\$
where $\widehat{f}_{\text{opt}}$ and $\widehat{B}$ is the maximizer of $V_{n,f}(B,d)$ for given $d$. By some heuristic analysis, as the working structural dimension $d = d_0$, we would have $V_{n,\widehat{f}_{\text{opt}}}^{(1)}(\widehat{B},d)$ converges to $0$ in probability, as $n \rightarrow \infty$; when $d > d_0$, we have $\text{IC}(d) > \text{IC}(d_0)$ with probability approaching to $1$; when $d < d_0$, we have $\text{IC}(d) = cn +  pd \log(n)$, where $c$ is some positive constant, and this is also larger than $\text{IC}(d_0)$ with probability approaching to $1$. Further theoretical analysis maybe required. However, we feel that this is beyond the scope of this paper.

\bibliographystyle{asa}
\bibliography{temp}

\end{document}